\documentclass[reprint, aps, prr, amsmath, amssymb, superscriptaddress]{revtex4-2}
\usepackage[utf8]{inputenc}
\usepackage{physics}
\usepackage{bm} 
\usepackage{graphicx} 
\usepackage{tikz} 
\usepackage{float}
\usetikzlibrary{arrows.meta, calc, positioning}
\usepackage[colorlinks=true, linkcolor=blue, citecolor=blue, urlcolor=blue]{hyperref}

\begin{document}
	
	\title{Tunable Statistics-Induced Caging in the Anyon-Hubbard Model}
	
	\author{Zenong Zhou}
	\affiliation{Zhejiang Key Laboratory of Quantum State Control and Optical Field Manipulation, Department of Physics, Zhejiang Sci-Tech University, Hangzhou 310018, China}
	
	\author{Chaorong Guo}
	\affiliation{Zhejiang Key Laboratory of Quantum State Control and Optical Field Manipulation, Department of Physics, Zhejiang Sci-Tech University, Hangzhou 310018, China}
	
	\author{Hongzheng Wu}
	\affiliation{Zhejiang Key Laboratory of Quantum State Control and Optical Field Manipulation, Department of Physics, Zhejiang Sci-Tech University, Hangzhou 310018, China}
	
	\author{Qianglin Hu}
	\affiliation{Department of Physics and Electronics Engineering, Tongren University, Tong Ren 554300, China}
	
	\author{Xiaobing Luo}
	\email{xiaobingluo2013@aliyun.com}
	\affiliation{Zhejiang Key Laboratory of Quantum State Control and Optical Field Manipulation, Department of Physics, Zhejiang Sci-Tech University, Hangzhou 310018, China}
	
	\date{\today}
	
	\begin{abstract}
		We study the quantum dynamics of two interacting anyons in the Anyon-Hubbard model on a four-site plaquette, a system that is exactly mappable to a Bose-Hubbard model. We reveal that static Aharonov-Bohm (AB) caging, induced by only specific statistical phases, emerges in the strongly interacting limit but breaks down under weak interparticle interactions. To address this, we demonstrate that statistical-factor-induced AB caging can be dynamically restored via Floquet engineering. This dynamical mechanism, governed by the synthetic Floquet flux and the anyonic statistical phase, extends the caging effect into the weakly interacting regime across the full spectrum of statistical phases. Furthermore, we show that the external drive enables the selective caging of anyons, providing an efficient approach for manipulating anyons and identifying statistical phases.
	\end{abstract}
	
	\maketitle
		
	\section{Introduction}
	Anyons are quantum quasiparticles exhibiting fractional exchange statistics intermediate between conventional bosons and fermions \cite{Leinaas1977, Wilczek1982, Halperin1984, Arovas1984, Stern2008, Nayak2008, Feldman2021, Trung2021}. Although the concept of anyons was established in the 1980s, the direct experimental observation of their fractional exchange statistics has only been achieved recently \cite{Bartolomei2020, Nakamura2020}. Today, anyons are fundamental to macroscopic phenomena such as fractional quantum Hall systems \cite{Stern2008} and topological quantum computation \cite{Nayak2008, Google2023}. 
	Parallel to these solid-state advances, one-dimensional anyons have been realized experimentally in ultracold atomic systems via density-dependent hopping engineered through Floquet modulation \cite{Strater2016, Schweizer2019, Kwan2024} and Raman-assisted tunneling \cite{Keilmann2011}, as well as in electric circuit simulators \cite{Zhang2022}. The anyonic statistical parameter has been shown to induce various dynamical effects, such as asymmetric quantum walks \cite{Kwan2024}, modified Bloch oscillations \cite{Longhi2012}, quantum phase transitions \cite{Greschner2015}, and doublon formation \cite{Cardarelli2016}. 
	Furthermore, the interplay between fractional statistics and advanced lattice engineering has recently unveiled highly exotic phenomena beyond conventional closed systems, such as the emergence of anyonic bound states in the continuum (BICs) \cite{Zhang2023BIC} and the dynamical suppression of the many-body non-Hermitian skin effect \cite{Qin2025}. In such interacting anyonic systems, the statistical phase explicitly dictates the underlying many-body quantum interference \cite{Chamon1997, Campagnano2012}.
	
	Quantum interference in discrete lattices, governed by synthetic gauge fields, provides a mechanism for manipulating particle dynamics \cite{Leykam2018}. In a minimal closed geometric unit, such as a four-site plaquette, a synthetic magnetic flux can induce destructive interference between clockwise and counterclockwise hopping paths \cite{Vidal1998}. When extended to a one-dimensional (1D) rhombus chain, this geometrical interference yields flat energy bands and suppresses single-particle transport. This static localization, termed Aharonov-Bohm (AB) caging, has been observed in diverse platforms including photonic lattices \cite{Mukherjee2018} and ultracold atoms \cite{Li2022}. Owing to its geometric origin, this single-particle caging is robust against uncorrelated static disorder \cite{Longhi2021}. However, in the many-body regime, the fate of this static caging depends on the interparticle interaction strength \cite{Doucot2002, Creffield2010}. While strong interactions can bind particles into composite doublons to recover localized states, the AB cage remains fragile against weak-to-moderate interactions \cite{Zurita2020}. In this weakly interacting regime, insufficient binding energy leads to multiparticle scattering, which disrupts the destructive interference conditions \cite{Zurita2020}. Consequently, this scattering-induced process breaks down the static caging, resulting in system-wide delocalization \cite{Vidal2000, Zurita2020}.
	
	To circumvent this scattering-induced delocalization, Floquet engineering via high-frequency driving offers a robust method to control quantum dynamics \cite{Bukov2015, Eckardt2017}. Time-periodic fields coherently renormalize hopping matrices, yielding tunable effective Hamiltonians \cite{Goldman2014, Eckardt2015} that can engineer phenomena like synthetic spin-orbit coupling and correlated tunneling \cite{Luo2021, Wu2024}. Furthermore, two-dimensional driving protocols enable simultaneous control over hopping amplitudes and synthetic gauge phases; the spatial geometry or polarization of such drives serves as an additional degree of freedom \cite{Struck2012, Jotzu2014}. Prominent quantum phenomena arising in such driven systems are coherent destruction of tunneling (CDT) and dynamic localization, characterized by the complete suppression of bare tunneling \cite{Grossmann1991, Lignier2007, Eckardt2009}. By selectively freezing transport channels and tuning the effective interaction-to-hopping ratio \cite{Zenesini2009}, this mechanism counteracts many-body scattering, successfully restoring localization in interacting systems \cite{Creffield2010}. However, standard CDT provides a strictly statistics-independent localization mechanism restricted to isolated driving parameter points. This limitation raises two key questions: first, whether the intrinsic statistical phase alone can induce AB caging for two interacting anyons; and second, whether a dynamical drive can coherently control this AB caging to realize the caging of anyons with arbitrary statistical parameters.
	
	In this work, we investigate the two-anyon dynamics of the Anyon-Hubbard model (AHM) with a four-site plaquette. In the undriven case, the statistical factor cannot induce AB caging---a phenomenon strictly preventing anyons initialized at site 1 from reaching the opposite diagonal site 3---within the weakly interacting regime, whereas certain specific statistical factors can induce caging in the strongly interacting regime. We demonstrate that a two-dimensional high-frequency drive lifts these restrictions, realizing statistical-factor-induced AB caging even in the weakly interacting regime. This dynamical caging is governed by an effective magnetic flux determined by both the intrinsic anyonic statistical phase and the drive-induced phase. Distinct from exact CDT, this flux-controlled caging persists over a broad range of driving strengths. Furthermore, by tuning the driving polarization angle, we can achieve caging for anyons with different statistical parameters, thereby enabling the identification of statistical factors.
	
	The rest of this paper is organized as follows. In Sec.~\ref{sec:model}, we present the Anyon-Hubbard model defined on a four-site plaquette and map it exactly to an equivalent bosonic Hamiltonian. Section~\ref{sec:static} is devoted to the undriven scenario of two interacting anyons, where we discover that AB caging is induced by a specific anyonic statistical phase in the strongly interacting regime. In Sec.~\ref{sec:driven}, we turn to the driven case, revealing that Floquet engineering can lead to AB caging for arbitrary statistical phases, even in the weakly interacting regime. By applying a perturbative framework to the effective Floquet Hamiltonian, we derive the analytical conditions for caging and further reveal a statistics-selective effect that can be tuned by the polarization angle of the drive. Finally, we conclude the paper with a summary of our key findings in Sec.~\ref{sec:conclusion}.

	\section{Model}
	\label{sec:model}
	
\begin{figure}[t!]
		\centering
		\resizebox{\columnwidth}{!}{
			\begin{tikzpicture}[
				scale=1.0, 
				site/.style={circle, draw=black, fill=white, inner sep=2pt, minimum size=24pt, ultra thick, font=\large}, 
				dim/.style={font=\small}
				]
				
				\coordinate (O) at (0,0);
				\coordinate (top) at (0,2.2);
				\coordinate (left) at (-2.2,0);
				\coordinate (bottom) at (0,-2.2);
				\coordinate (right) at (2.2,0);
				
				\draw[gray, dashed, thick] (left) -- (O);
				\draw[gray, dashed, thick] (top) -- (O);
				\draw[gray, dashed, thick] (bottom) -- (O);
				\draw[gray, dashed, thick] (O) -- node[below, text=black, font=\Large, inner sep=5pt] {$a$} (right);
				\filldraw[gray] (O) circle (2pt); 
				
				\draw[line width=2.5pt, black] (left) -- node[above left=2pt, font=\large] {$J$} (top); 
				\draw[line width=2.5pt, black] (top) -- node[above right=2pt, font=\large] {$J$} (right); 
				\draw[line width=2.5pt, black] (right) -- node[below right=2pt, font=\large] {$J$} (bottom); 
				
				\foreach \i in {0,1,...,100} {
					\pgfmathsetmacro{\hue}{\i/125} 
					\pgfmathsetmacro{\posA}{\i/100}
					\pgfmathsetmacro{\posB}{(\i+1.5)/100} 
					\definecolor{rainbowcolor}{hsb}{\hue, 1, 1}
					\draw[line width=2.5pt, rainbowcolor] ($(bottom)!\posA!(left)$) -- ($(bottom)!\posB!(left)$);
				}
				\node[below left=2pt, text=black, font=\large] at ($(bottom)!0.5!(left)$) {$J$};
				
				\node[site] (1) at (left) {$1$};
				\node[site] (2) at (top) {$2$};
				\node[site] (3) at (right) {$3$};
				\node[site] (4) at (bottom) {$4$};
				
				\draw[blue, line width=3.5pt, ->, >=Stealth] ($(top)+(-1.4,1.4)$) 
				node[above left, font=\LARGE\bfseries, inner sep=1pt] {$\Delta_2$} 
				-- ($(top)+(-0.5,0.5)$);
				
				\draw[blue, line width=3.5pt, ->, >=Stealth] ($(bottom)+(1.4,-1.4)$) 
				node[below right, font=\LARGE\bfseries, inner sep=1pt] {$\Delta_4$} 
				-- ($(bottom)+(0.5,-0.5)$);
				
				\begin{scope}[shift={(4.6, 1.2)}, scale=1.0] 
					\draw[->, >=Stealth, thick] (0,0) -- (2.3,0) node[right, font=\large] {$x$};
					\draw[->, >=Stealth, thick] (0,0) -- (0,2.3) node[above, font=\large] {$y$};
					\draw[->, >=Stealth, ultra thick, red] (0,0) -- (1.5, 1) node[right, text=black, font=\large] {$\vec{F}(t)$};
					\draw[gray, dashed, thick] (1.5,0) |- (0,1);
					
					\draw[->, gray, thick] (0.8,0) arc (0:33.6:0.8);
					\node[font=\large] at (1.1, 0.35) {$\theta$};
					
					\node[right, font=\large] at (-0.6, -0.8) {$F_x(t) = A\cos\theta\cos(\omega t)$};
					\node[right, font=\large] at (-0.6, -1.5) {$F_y(t) = A\sin\theta\sin(\omega t)$};
				\end{scope}
				
				\begin{scope}[shift={(-3.4, 2.2)}, scale=0.6] 
					\draw[thick] (0,0) ellipse (0.8cm and 0.5cm);
					\filldraw[gray] (-0.35,0.2) circle (0.18cm);    
					\filldraw[black] (0.25, -0.15) circle (0.18cm); 
					\node[above, font=\large\bfseries] at (-0.3,0.7) {Anyon};
				\end{scope}
			\end{tikzpicture}
		} 
		\caption{Schematic of the diamond-shaped four-site plaquette with nearest-neighbor hopping amplitude $J$ and center-to-vertex distance $a$. The top-left inset illustrates the two anyons. The link between sites 1 and 4, highlighted with a color gradient, represents the twisted boundary condition induced by the anyon-to-boson mapping. Blue arrows indicate the static energy detunings $\Delta_2$ and $\Delta_4$. The right inset defines the driving field $\vec{F}(t)$ with amplitude $A$, frequency $\omega$, and polarization angle $\theta$.}
		\label{fig:1}
	\end{figure}
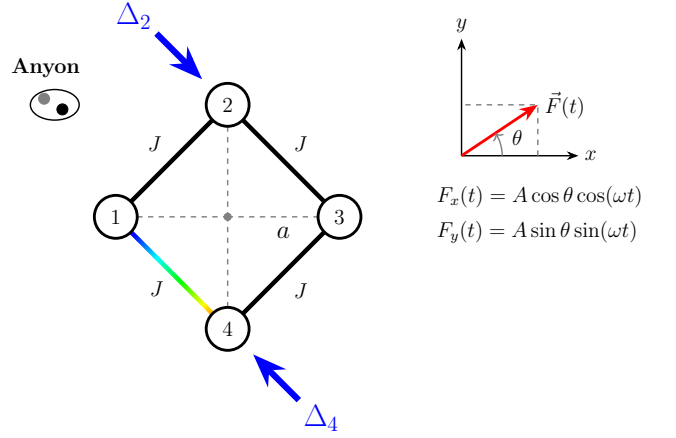

	We consider a diamond-shaped four-site plaquette, as depicted in Fig.~\ref{fig:1}. The system dynamics, governed by constrained tunneling and external Floquet driving, are described by the periodically driven Anyon-Hubbard model. The Hamiltonian is given by:
	\begin{equation} \label{eq:H_AHM}
		\begin{split}
			\hat{H}_{\text{AHM}}(t) = &-J \sum_{j=1}^3 \left( \hat{a}_j^\dagger \hat{a}_{j+1} + \text{H.c.} \right) - J \left( \hat{a}_4^\dagger \hat{a}_1 + \text{H.c.} \right) \\
			&+ \frac{U}{2} \sum_{j=1}^4 \hat{n}_j(\hat{n}_j - 1) + \sum_{j=1}^4 V_j(t) \hat{n}_j.
		\end{split}
	\end{equation}
	
	Here, $\hat{a}_j^\dagger$ ($\hat{a}_j$) represents the anyonic creation (annihilation) operator at site $j$, which obeys the fractional commutation relations $\hat{a}_j \hat{a}_k^\dagger - \mathrm{e}^{-i\kappa\pi\text{sgn}(j-k)} \hat{a}_k^\dagger \hat{a}_j = \delta_{jk}$, where the statistical parameter $\kappa \in [0, 1]$ interpolates between bosonic ($\kappa = 0$) and pseudo-fermionic ($\kappa = 1$) statistics. The parameters $J$ and $U$ represent the hopping amplitude and on-site interaction strength, respectively. The scalar potential $V_j(t) \equiv \vec{F}(t) \cdot \vec{r}_j + \Delta_j$ incorporates the external high-frequency driving field $\vec{F}(t)$ and static detunings $\Delta_j$.
	
	To project the fractional statistics onto a standard Fock space, we utilize the fractional Jordan-Wigner transformation \cite{Keilmann2011, Greschner2014}:
	\begin{equation} \label{eq:JWT}
		\hat{a}_j = \hat{\tilde{b}}_j \exp\left( i\kappa\pi \sum_{k<j} \hat{n}_k \right).
	\end{equation}
	The mapping leaves the local density operator invariant ($\hat{n}_j = \hat{a}_j^\dagger \hat{a}_j = \hat{\tilde{b}}_j^\dagger \hat{\tilde{b}}_j$); thus, the on-site interaction and scalar potential terms retain their original form. Upon applying this transformation, the Anyon-Hubbard Hamiltonian maps to an  bosonic Hamiltonian $\hat{\tilde{H}}^B_{\text{AHM}}$, which decomposes into a bulk sum and a boundary hopping term that encodes the twisted boundary condition:
	\begin{equation} \label{eq:H_intermediate}
		\begin{split}
			\hat{\tilde{H}}^B_{\text{AHM}}(t) &= -J \sum_{j=1}^3 \left( \hat{\tilde{b}}_j^\dagger \hat{\tilde{b}}_{j+1} e^{i\kappa\pi\hat{n}_j} + \text{H.c.} \right) \\
			&\quad - J \left( \hat{\tilde{b}}_4^\dagger \hat{\tilde{b}}_1 e^{i\kappa\pi\hat{n}_4} e^{-i\kappa\pi(N-1)} + \text{H.c.} \right) \\
			&\quad + \frac{U}{2} \sum_{j=1}^4 \hat{n}_j(\hat{n}_j - 1) + \sum_{j=1}^4 V_j(t)\hat{n}_j.
		\end{split}
	\end{equation}
	The boundary hopping term can be expressed as:
	\begin{equation} \label{eq:boundary_hopping}
		\hat{a}_4^\dagger \hat{a}_1 = \hat{\tilde{b}}_4^\dagger e^{-i\kappa\pi(\hat{N}-\hat{n}_4)} \hat{\tilde{b}}_1 = \hat{\tilde{b}}_4^\dagger \hat{\tilde{b}}_1 e^{i\kappa\pi\hat{n}_4} e^{-i\kappa\pi(N-1)}.
	\end{equation}
	To obtain this result, we simplify the non-local phase string using the relation $\sum_{k=1}^3 \hat{n}_k = \hat{N} - \hat{n}_4$. Commuting $\hat{\tilde{b}}_1$ through this phase string utilizes the relation $e^{-i\kappa\pi\hat{N}} \hat{\tilde{b}}_1 = \hat{\tilde{b}}_1 e^{-i\kappa\pi(\hat{N}-1)}$, followed by replacing the operator $\hat{N}$ with its eigenvalue $N$.
	
	Although the formalism holds for an arbitrary particle number $N$, we now focus on the two-particle sector ($N = 2$) to make the discussion concrete. In this sector, the residual boundary phase simplifies to $e^{-i\kappa\pi}$. To eliminate this spatial asymmetry and distribute the statistical phase uniformly around the loop, we apply the unitary gauge transformation $\hat{\mathcal{U}} = \exp\left[ i \sum_{m=1}^4 m\frac{\kappa\pi}{4} \hat{n}_m \right]$. The canonical bosonic operators are then defined as:
	\begin{equation}\label{eq:gauge_transform}
		\hat{b}_j = \hat{\mathcal{U}}^\dagger \hat{\tilde{b}}_j \hat{\mathcal{U}} = \hat{\tilde{b}}_j e^{ij\frac{\kappa\pi}{4}}.
	\end{equation}
	Substituting the inverse relation $\hat{\tilde{b}}_j = \hat{b}_j e^{-ij\frac{\kappa\pi}{4}}$ into Eq.~\eqref{eq:H_intermediate} causes the bulk and boundary terms to recombine, yielding the translationally invariant form of the full bosonic Hamiltonian $\hat{H}_{\text{AHM}}^B(t)$:
	\begin{equation}\label{eq:full_hamiltonian}
		\begin{aligned}
			\hat{H}_{\text{AHM}}^B(t) &= \hat{H}_{\text{static}} + \sum_{j=1}^4 V_j(t)\hat{n}_j \\
			&= -J \sum_{j=1}^4 \left( \hat{b}_j^\dagger \hat{b}_{j+1} e^{i\kappa\pi\hat{n}_j} e^{-i\frac{\kappa\pi}{4}} + \text{H.c.} \right) \\
			&\quad + \frac{U}{2} \sum_{j=1}^4 \hat{n}_j(\hat{n}_j - 1) + \sum_{j=1}^4 V_j(t)\hat{n}_j,
		\end{aligned}
	\end{equation}
	with the periodic boundary condition $\hat{b}_5 \equiv \hat{b}_1$. Compared to the bosonic representation of the standard anyon-Hubbard model (which typically corresponds to open boundary conditions), the uniform phase factor $e^{-i\kappa\pi/4}$ encodes the anyonic boundary twist distributed evenly across the hopping links, thereby restoring translational invariance via the local unitary transformation. Experimentally, anyonic statistical phases of this type—appearing as density-dependent hopping amplitudes—have been synthesized in ultracold atomic platforms via, for example, Raman-assisted tunneling techniques \cite{Keilmann2011}, which enable the statistical angle to be controlled in situ by modifying the relative phase of the Raman fields.
	
	\section{The Undriven case}
	\label{sec:static}
	
	We investigate the dynamics of two anyons on a four-site plaquette by initially setting external modulations to zero ($V_j(t) = 0$). The coherent evolution of this static system is governed by the dimensionless Schrödinger equation:
	\begin{equation} \label{eq:Schrodinger}
		i\frac{\partial}{\partial t}|\psi(t)\rangle = \hat{H}_{\text{static}}|\psi(t)\rangle.
	\end{equation} 
	
	In the following analysis, we adopt the hopping amplitude $J$ as the unit of energy; consequently, time is measured in units of $\hbar/J$ and the interaction strength is parameterized by $U/J$. The Anyon-Hubbard model can be experimentally realized using ultracold atoms, with typical energy scales of $J/\hbar \sim 0.1\text{--}1\text{ kHz}$ \cite{Keilmann2011, Lignier2007}, and $U$ tunable up to several tens of $J$ \cite{Greschner2014, Schweizer2019}.
	
	Expanding the state vector $|\psi(t)\rangle$ in the complete real-space Fock basis, the basis states $|i, j\rangle$ for the two-particle manifold ($\hat{N} = 2$) are defined as:
	\begin{equation} \label{eq:basis}
		|i, j\rangle = \frac{1}{\sqrt{1 + \delta_{i,j}}} \hat{b}_i^\dagger \hat{b}_j^\dagger |0\rangle, \quad (1 \leq i \leq j \leq 4)
	\end{equation}
	where $|0\rangle$ denotes the vacuum state. The ordered index notation ($i \leq j$) encompasses the six states with particles on distinct sites ($i < j$) as well as the four localized doublon states ($i = j$). Projecting the dynamics onto this ten-dimensional Hilbert space, the general wave function is expressed as $|\psi(t)\rangle = \sum_{1 \leq i \leq j \leq 4} c_{i,j}(t)|i, j\rangle$. Substituting this expansion into Eq.~\eqref{eq:Schrodinger} yields a closed set of coupled ordinary differential equations for the complex probability amplitudes $c_{i,j}(t)$.
	
	In order to examine the caging effect, we prepare the system in the state $|\psi(0)\rangle = |1,1\rangle$, where both particles are initially localized at the injection site ($j = 1$). We focus on the time-dependent particle occupation at the opposite site ($j = 3$), given by the expectation value $\langle\hat{n}_3(t)\rangle = \langle\psi(t)|\hat{b}_3^\dagger \hat{b}_3|\psi(t)\rangle$. To average out coherent oscillations and extract the steady-state characteristics, we define the long-time averaged occupation as
	\begin{equation} \label{eq:time_average}
		\overline{\langle\hat{n}_3\rangle} = \frac{1}{\tau} \int_0^\tau \langle\psi(t)|\hat{b}_3^\dagger \hat{b}_3|\psi(t)\rangle dt,
	\end{equation}
	for a sufficiently long evolution time $\tau$.	
	
\begin{figure}[htbp]
		\centering
		\includegraphics[width=\linewidth]{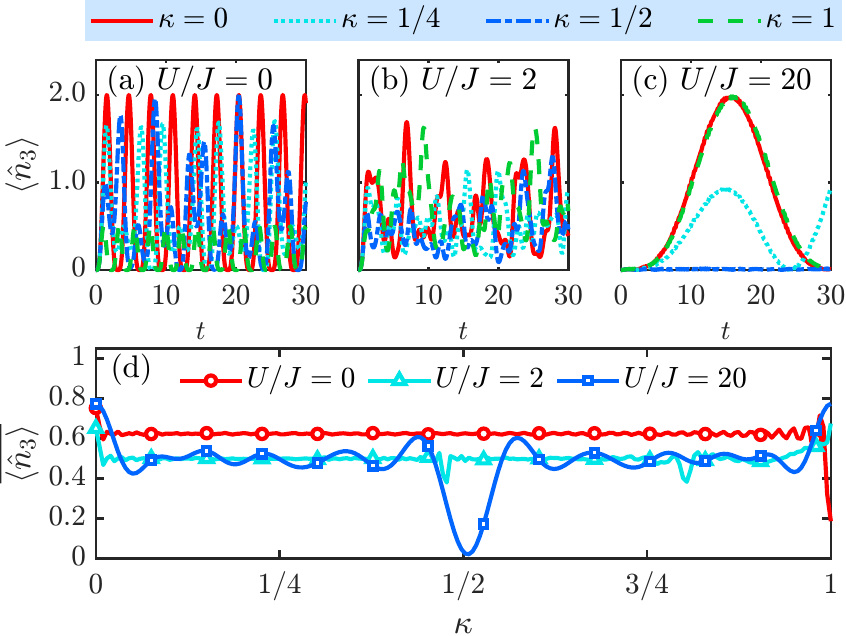}
		\caption{Quantum dynamics of the target site occupation $\langle\hat{n}_3(t)\rangle$ governed by Eq.~\eqref{eq:Schrodinger}. (a)–(c) Real-time evolution for various statistical parameters $\kappa$ at interaction strengths $U/J=0$, $2$, and $20$, respectively. (d) Long-time-averaged occupation $\overline{\langle \hat{n}_3 \rangle}$ as a function of $\kappa$. Here and in all subsequent figures, time $t$ is expressed in units of $\hbar/J$, and the system is initially prepared with both particles localized at site 1, denoted as $|\psi(0)\rangle = |1,1\rangle$. }
		\label{fig:2}
	\end{figure}

	Figure~\ref{fig:2} illustrates the quantum dynamics of the target site occupation $\langle\hat{n}_3(t)\rangle$, where the interaction strength $U/J$ dictates the dynamical regimes. For weak interactions ($U/J = 0, 2$), the target site occupation exhibits oscillatory behavior over time for different statistical parameters, as shown in Figs.~\ref{fig:2}(a) and (b). In contrast, under strong interaction ($U/J = 20$), a statistically induced caging is observed in Fig.~\ref{fig:2}(c): when the statistical parameter $\kappa = 1/2$, the occupation of the target site is completely suppressed ($\langle\hat{n}_3(t)\rangle \equiv 0$), while the curves for other $\kappa$ ($\kappa=0, 1/4, 1$) continue to oscillate.
	Fig.~\ref{fig:2}(d) shows the long-time averaged occupation $\overline{\langle\hat{n}_3\rangle}$ as a function of $\kappa$ for different interaction strengths. We observe that for weak interactions ($U/J=0, 2$), $\overline{\langle\hat{n}_3\rangle}$ exhibits finite values as $\kappa$ varies, indicating no caging effect. However, for strong interaction ($U/J=20$), when $\kappa = 1/2$, $\overline{\langle\hat{n}_3\rangle}$ drops to zero, which clearly demonstrates the caging phenomenon (i.e., the tunneling to the target site is suppressed).
	
\begin{figure}[t!]
		\centering
		\includegraphics[width=\linewidth]{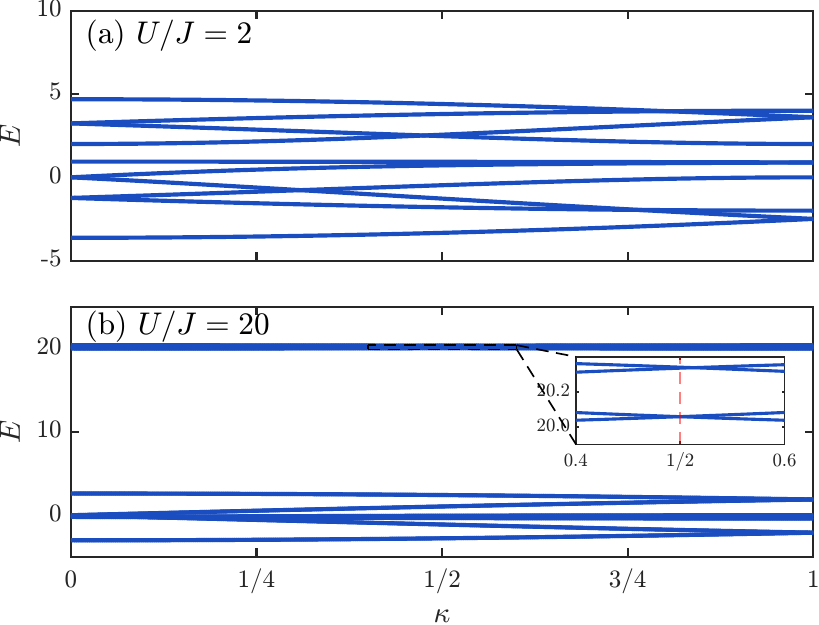}
		\caption{Energy spectra of the undriven two-anyon system governed by $\hat{H}_{\text{static}}$ as a function of the statistical parameter $\kappa$ for interaction strengths (a) $U/J = 2$ and (b) $U/J = 20$. In (b), the energy levels separate into a lower scattering manifold and an upper doublon band. The inset resolves the fine structure of the doublon band, revealing an exact degeneracy at $\kappa = 1/2$ (red dashed line).}
		\label{fig:3}
	\end{figure}

	To elucidate the statistical-parameter-induced caging effect in the strongly interacting regime (Fig.~\ref{fig:2}), we calculate the energy spectrum of the undriven two-anyon system, governed by $\hat{H}_{\text{static}}$, as a function of $\kappa$ for $U/J = 2$ and $20$. In contrast to the weakly interacting regime, which is dominated by scattering (unpaired) states [Fig.~\ref{fig:3}(a)], strong interactions split the spectrum into two distinct bands: a broad lower band of unpaired states (with particles on distinct sites) and a narrow upper band of doublon states (with both particles on the same site) [Fig.~\ref{fig:3}(b)]. As highlighted in the inset, the doublon levels exhibit a twofold degeneracy precisely at $\kappa = 1/2$, coinciding with the onset of the caging effect. This degeneracy induces exact destructive quantum interference between the clockwise and counterclockwise tunneling paths of doublon, directly leading to the suppressed tunneling observed in Fig.~\ref{fig:2} (analyzed in detail below). Consistent with this mechanism, an examination of the probability amplitudes $c_{i,j}(t)$ from the numerical data in Fig.~\ref{fig:2}(c) confirms that $c_{i,j}(t) \approx 0$ for all $i \neq j$ throughout the evolution (data not shown for brevity). This verifies that two strongly interacting particles initially localized on the same site form a bound doublon, which undergoes correlated tunneling across the four-site lattice.

	We employ second-order degenerate perturbation theory in the strongly interacting regime ($U \gg J$) to reveal the physical origin of this degeneracy and the resulting caging effect. We decompose the undriven Hamiltonian into an unperturbed interaction term and a hopping perturbation, defined as $\hat{H}_{\text{static}} = \hat{H}_U + \hat{H}_{\text{hop}}$. Here, the interaction term $\hat{H}_U = \frac{U}{2} \sum_{j=1}^4 \hat{n}_j(\hat{n}_j - 1)$ dominates, while the nearest-neighbor hopping term $\hat{H}_{\text{hop}} = -J \sum_{j=1}^4 \left( \hat{b}_j^\dagger \hat{b}_{j+1} e^{-i\frac{\kappa\pi}{4}} e^{i\kappa\pi\hat{n}_j} + \text{H.c.} \right)$ is treated as a small perturbation.
	We focus on the doublon subspace spanned by the four localized states ${|2\rangle_j}$, where $|2\rangle_j$ denotes two particles occupying site $j$. In the Fock basis $|i, j\rangle$ defined in Eq.~\eqref{eq:basis}, these doublon states are expressed as $|2\rangle_j \equiv |j, j\rangle = \frac{1}{\sqrt{2}}(\hat{b}_j^\dagger)^2|0\rangle$. The remaining six states $|i, j\rangle$ (with $i < j$) represent separated particles and constitute the orthogonal manifold satisfying $\hat{H}_U|i, j\rangle = 0$.
	The effective Hamiltonian projected onto the doublon subspace is expanded perturbatively to second order:
	\begin{equation} \label{eq:H_eff}
		\hat{H}_{\text{eff}}^D = \hat{h}^{(0)} + \hat{h}^{(1)} + \hat{h}^{(2)}.
	\end{equation}
	The zeroth-order term corresponds to the unperturbed interaction energy of the localized doublons:
	\begin{equation} \label{eq:h_0}
		\hat{h}^{(0)} = U \sum_{j=1}^4 |2\rangle_j \langle 2|_j.
	\end{equation}
	The first-order correction vanishes because a single hopping event cannot directly connect two doublon states:
	\begin{equation} \label{eq:h_1}
		\hat{h}^{(1)} = \sum_{j,k} \langle 2|_k \hat{H}_{\text{hop}} |2\rangle_j |2\rangle_k \langle 2|_j = 0.
	\end{equation}
	The second-order term captures the virtual dissociation of a doublon into the zero-energy manifold of separated (unpaired) particles, followed by its recombination. Since $\hat{H}_{\text{hop}}$ moves only one particle to an adjacent site when acting on a localized doublon $|2\rangle_j$, the relevant intermediate states are restricted to nearest-neighbor separated pairs (specifically, $|j, j+1\rangle$ or $|j-1, j\rangle$ under periodic boundary conditions). Evaluating this perturbation series yields:
	\begin{equation} \label{eq:h2}
		\begin{split}
			\hat{h}^{(2)} &= \sum_{j,k} \sum_{i'<j'} \frac{\langle 2|_k \hat{H}_{\text{hop}} |i', j'\rangle \langle i', j'| \hat{H}_{\text{hop}} |2\rangle_j}{U} |2\rangle_k \langle 2|_j \\
			&= \sum_{j=1}^4 \frac{4J^2}{U} |2\rangle_j \langle 2|_j + \sum_{j=1}^4 \left( \frac{2J^2}{U} e^{-i\frac{\kappa\pi}{2}} |2\rangle_{j+1} \langle 2|_j + \text{H.c.} \right).
		\end{split}
	\end{equation}
	In this expression, the first summation represents a uniform on-site energy shift of $4J^2/U$, contributing a constant baseline to the energy spectrum. The second summation describes effective nearest-neighbor hopping of doublons, which acquires an induced phase of $-\kappa\pi/2$ per hop.
	
	By summing the zeroth-order and second-order terms (noting that $\hat{h}^{(1)} = 0$), we obtain the full effective Hamiltonian within the doublon subspace:
	\begin{equation} \label{eq:H_eff_final}
		\begin{split}
			\hat{H}_{\text{eff}}^D &= \left( U + \frac{4J^2}{U} \right) \sum_{j=1}^4 |2\rangle_j \langle 2|_j \\
			&\quad + \sum_{j=1}^4 \left( \frac{2J^2}{U} e^{-i\phi} |2\rangle_{j+1} \langle 2|_j + \text{H.c.} \right),
		\end{split}
	\end{equation}
	where $\phi \equiv \kappa\pi/2$ denotes the effective hopping phase induced by the fractional statistics, and we impose the periodic boundary condition $|2\rangle_5 \equiv |2\rangle_1$.
	
	To solve the stationary Schrödinger equation $\hat{H}_{\text{eff}}^D|\psi\rangle = E|\psi\rangle$, we expand the eigenstate in the doublon basis as $|\psi\rangle = \sum_{j=1}^4 c_{j}|2\rangle_j$. Substituting this into the eigenvalue problem and defining the relative energy $\tilde{E} = E - \left(U + \frac{4J^2}{U}\right)$ yields the discrete difference equation:
	\begin{equation} \label{eq:difference_eq}
		\tilde{E}c_{j} = \frac{2J^2}{U} \left( e^{i\phi} c_{j+1} + e^{-i\phi} c_{j-1} \right), \quad (j = 1, 2, 3, 4)
	\end{equation}
	subject to the periodic boundary condition $c_{j+4} \equiv c_{j}$.
	
	Leveraging the alternating connectivity of the four-site loop, we eliminate the amplitudes on the even sites, we eliminate the amplitudes on the even sites ($c_{2}$ and $c_{4}$). This reduction of the $4 \times 4$ linear system leads to the following $2 \times 2$ secular equation for the amplitudes on the odd sites ($c_{1}$ and $c_{3}$):
	\begin{equation} \label{eq:secular_matrix}
		\left\{
		\begin{aligned}
			&\left( \tilde{E}^2 - \frac{8J^4}{U^2} \right) c_{1} - \frac{8J^4}{U^2}\cos(2\phi)c_{3} = 0, \\
			&-\frac{8J^4}{U^2}\cos(2\phi)c_{1} + \left( \tilde{E}^2 - \frac{8J^4}{U^2} \right) c_{3} = 0.
		\end{aligned}
		\right.
	\end{equation}

	The effective coupling between sites 1 and 3 is governed by the off-diagonal term proportional to $\cos(2\phi)$. To suppress tunneling to the target site, this coupling must vanish, yielding the condition:
	\begin{equation} \label{eq:static_caging}
		\cos(2\phi) = 0 \implies \kappa = \frac{1}{2}.
	\end{equation}
	Physically, as a doublon tunnels around the four-site plaquette, it accumulates a total phase $\Phi = 4\phi = 2\kappa\pi$. The condition $\kappa = 1/2$ corresponds to a flux of $\Phi = \pi$, which results in complete destructive interference; this localization effect is known as Aharonov-Bohm caging.
	
	Substituting this condition into Eq.~\eqref{eq:secular_matrix} eliminates the effective long-range hopping term. Solving the resulting decoupled equations gives $\tilde{E}^2 = \frac{8J^4}{U^2}$. Restoring the constant energy shift $U + \frac{4J^2}{U}$, the four doublon eigenenergies form two degenerate pairs:
	\begin{equation} \label{eq:eigenenergies}
		E_{1,2} = U + \frac{4J^2}{U} + \frac{2\sqrt{2}J^2}{U}, \quad E_{3,4} = U + \frac{4J^2}{U} - \frac{2\sqrt{2}J^2}{U}.
	\end{equation}

	This analytical result fully elucidates why the dynamical trapping in Fig.~\ref{fig:2} occurs specifically at $\kappa = 1/2$. Furthermore, it demonstrates that the twofold crossing of the doublon energy levels observed in Fig.~\ref{fig:3}(b) stems from this same underlying interference mechanism.

	\section{Floquet Engineering of Dynamical Caging}
	\label{sec:driven}
	
	We now extend our investigation to the driven regime. The time-dependent scalar potential at site $j$ is defined as $V_j(t) \equiv \vec{F}(t) \cdot \vec{r}_j + \Delta_j$, where the driving field $\vec{F}(t)$ in the $x$-$y$ plane is given by:
	\begin{equation} \label{eq:driving_field}
		\vec{F}(t) = A \cos\theta \cos(\omega t)\hat{e}_x + A \sin\theta \sin(\omega t)\hat{e}_y.
	\end{equation}
	Here, $A$ denotes the driving amplitude, $\omega = 2\pi/T$ is the angular frequency corresponding to the driving period $T$, and $\theta$ represents the polarization angle.
	
	The site coordinates of the diamond unit cell are $\vec{r}_1 = (-a, 0)$, $\vec{r}_2 = (0, a)$, $\vec{r}_3 = (a, 0)$, and $\vec{r}_4 = (0, -a)$. We introduce static energy detunings $\Delta_1 = \Delta_3 = 0$ and $\Delta_2 = \Delta_4 = m\omega$, where $m \in \mathbb{Z}$ is the photon resonance order. These detunings suppress the bare single-particle hopping by creating a large energy mismatch between neighboring sites. A high-frequency drive ($\omega \gg J/\hbar$) restores nearest-neighbor tunneling through the resonant absorption or emission of $m$ photons \cite{Lignier2007, Struck2012}. This photon-assisted tunneling imprints spatially dependent phases onto the effective hopping amplitudes, thereby breaking time-reversal symmetry \cite{Jotzu2014}. 
	
	Specifically, substituting these coordinates and detunings yields the time-dependent scalar potentials:
	\begin{equation} \label{eq:scalar_potentials}
		V_j(t) = \begin{cases}
			-Aa \cos\theta \cos(\omega t), & j = 1, \\
			Aa \sin\theta \sin(\omega t) + m\omega, & j = 2, \\
			Aa \cos\theta \cos(\omega t), & j = 3, \\
			-Aa \sin\theta \sin(\omega t) + m\omega, & j = 4.
		\end{cases}
	\end{equation}
	In typical optical lattice experiments, the lattice constant is $a \approx 0.4-0.5~\mu\text{m}$, which corresponds to a bare hopping rate $J/\hbar \sim 0.1\text{--}1\text{ kHz}$. For such parameters, the driving frequency is chosen to be $\omega/2\pi \sim 40\text{ kHz}$, which maintains the validity of the high-frequency approximation \cite{Eckardt2017, Jotzu2014}. 
	
	\begin{figure}[t!]
		\centering
		\includegraphics[width=\linewidth]{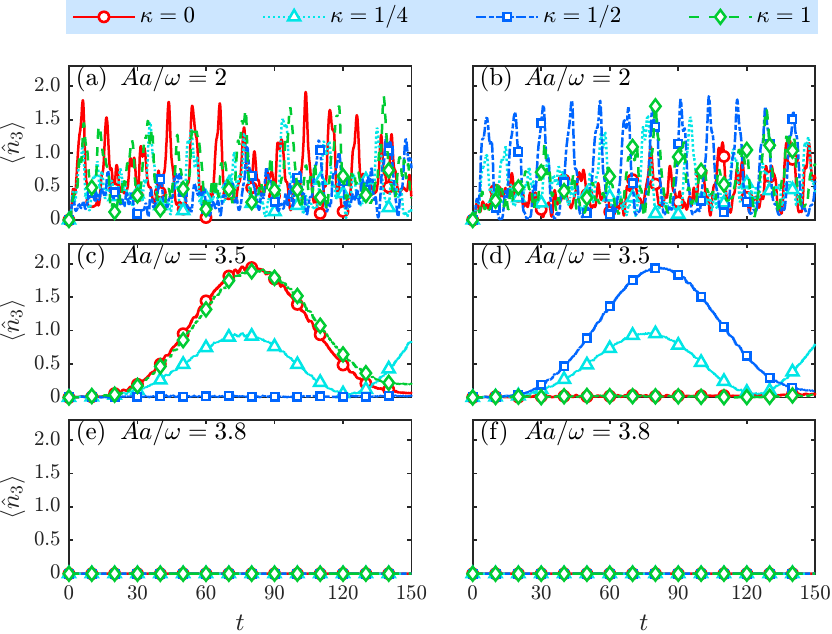} 
		\caption{Quantum dynamics of the target site occupation $\langle\hat{n}_3(t)\rangle$ governed by $\hat{H}_{\text{AHM}}^B(t)$ [Eqs.~\eqref{eq:full_hamiltonian} and \eqref{eq:scalar_potentials}]. The real-time evolution is calculated for an interaction strength $U/J = 2$, a driving frequency $\omega/J = 40$, and static detunings $\Delta_2 = \Delta_4 = \omega$. Results are displayed for dimensionless driving amplitudes $Aa/\omega = 2$ [(a), (b)], $3.5$ [(c), (d)], and $3.8$ [(e), (f)]. The left and right columns correspond to polarization angles $\theta = 0$ and $\theta = \pi/8$, respectively. Different curves within each panel represent various statistical parameters $\kappa$.}
		\label{fig:4}
	\end{figure}
	
	We investigate the influence of periodic driving on the caging effect by numerically solving the time-dependent Schrödinger equation, $i\frac{\partial}{\partial t}|\psi(t)\rangle = \hat{H}_{\text{AHM}}^B(t)|\psi(t)\rangle$, incorporating the scalar potentials defined in Eq.~\eqref{eq:scalar_potentials}. Adopting the initial state $|\psi(0)\rangle = |1,1\rangle$ and the observables from the static analysis, we calculate the target site occupation dynamics $\langle \hat{n}_3(t) \rangle$ for the weakly interacting regime ($U/J = 2$), as presented in Fig.~\ref{fig:4}. The figure systematically compares the time evolution for four distinct statistical parameters $\kappa$ under varying driving strengths ($Aa/\omega = 2, 3.5, 3.8$) and polarization angles ($\theta = 0$ and $\pi/8$ in the left and right columns, respectively).
	
	As established in the previous section, weak interactions ($U/J = 2$) fail to induce statistical caging in the undriven regime. Upon introducing high-frequency driving, the system dynamics exhibit a marked dependence on the driving parameters. At $Aa/\omega = 2$, the oscillatory behavior of $\langle \hat{n}_3(t) \rangle$ persists for all statistical parameters $\kappa$ at both $\theta = 0$ [Fig.~\ref{fig:4}(a)] and $\theta = \pi/8$ [Fig.~\ref{fig:4}(b)], analogous to the undriven scenario. However, at $Aa/\omega = 3.5$, the system dynamically restores the caging effect specifically at $\kappa = 1/2$, yielding $\langle \hat{n}_3(t) \rangle \equiv 0$ for $\theta = 0$ [Fig.~\ref{fig:4}(c)]. Shifting the polarization angle to $\theta = \pi/8$ [Fig.~\ref{fig:4}(d)] transfers this effect exclusively to the bosonic ($\kappa = 0$) and pseudo-fermionic ($\kappa = 1$) limits. Furthermore, at $Aa/\omega = 3.8$ [Figs.~\ref{fig:4}(e) and \ref{fig:4}(f)], which coincides with a zero of the first-order Bessel function, we observe a universal suppression of the target site occupation across all $\kappa$ and $\theta$. This points to a parameter-independent dynamical localization mechanism arising from the coherent destruction of tunneling (CDT).
	
	\begin{figure}[t!]
		\centering
		\includegraphics[width=\linewidth]{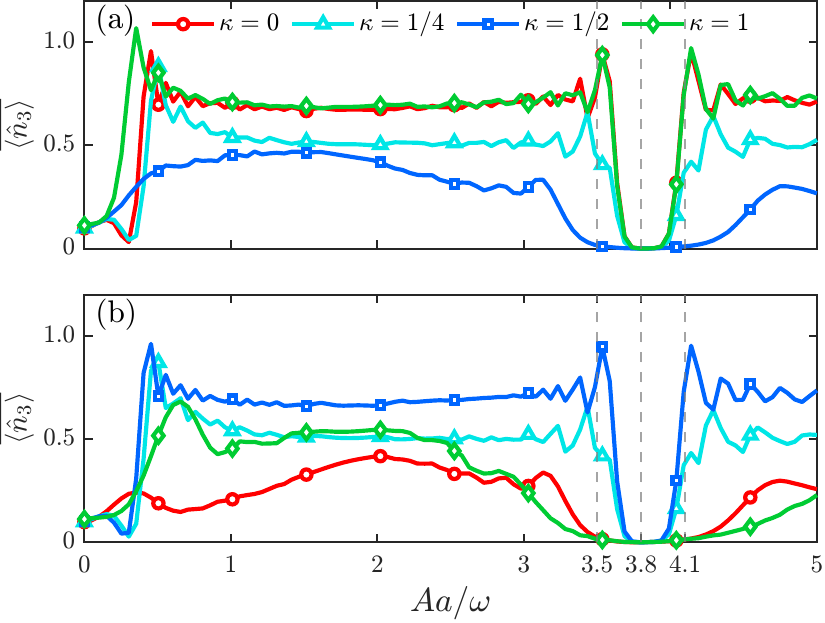}
		\caption{Long-time-averaged occupation $\overline{\langle \hat{n}_3 \rangle}$ governed by $\hat{H}_{\text{AHM}}^B(t)$ [Eqs.~\eqref{eq:full_hamiltonian} and \eqref{eq:scalar_potentials}] as a function of the dimensionless driving strength $Aa/\omega$ at $U/J = 2$ and $\omega/J = 40$. Results are shown for polarization angles (a) $\theta = 0$ and (b) $\theta = \pi/8$. The $\kappa$-dependent dips within the range $Aa/\omega \in [3.5, 4.1]$, marked by vertical dashed lines, signify dynamical caging, while the $\kappa$-independent dip at $Aa/\omega \approx 3.8$ corresponds to coherent destruction of tunneling (CDT).}
		\label{fig:5}
	\end{figure}
	
	The Floquet-engineered caging effect induced by statistical parameters is more clearly revealed by the long-time-averaged occupation of the target site, $\langle\hat{n}_3\rangle$, plotted versus the driving amplitude $Aa/\omega$ for polarization angles $\theta=0$ [Fig.~\ref{fig:5}(a)] and $\theta=\pi/8$ [Fig.~\ref{fig:5}(b)]. In both cases, the occupation is suppressed as $Aa/\omega\to0$ (the undriven limit) due to large-detuning localization, and it drops to zero exactly at $Aa/\omega=3.8$ for all four statistical parameters. This zero corresponds to the first root of the first-order Bessel function and signals coherent destruction of tunneling (CDT). Beyond this statistics-independent CDT, a statistics-dependent caging effect emerges. For $\theta=0$ [Fig.~\ref{fig:5}(a)], complete caging—$\langle\hat{n}_3\rangle=0$ throughout the interval $Aa/\omega\in[3.5,4.1]$—occurs only at $\kappa=1/2$. In contrast, for $\theta=\pi/8$ [Fig.~\ref{fig:5}(b)], such broad-range caging is observed solely in the bosonic ($\kappa=0$) and pseudo-fermionic ($\kappa=1$) limits. These findings indicate that tuning the driving polarization angle can selectively cage anyons carrying different statistical phases. Importantly, whereas CDT is intrinsically a narrow resonance (the small but finite width seen in the simulation is purely a numerical artifact caused by the finite evolution time), the statistics-selective caging demonstrated here persists over a wide range of driving amplitude.
	
\begin{figure}[t!]
		\centering
		\includegraphics[width=\linewidth]{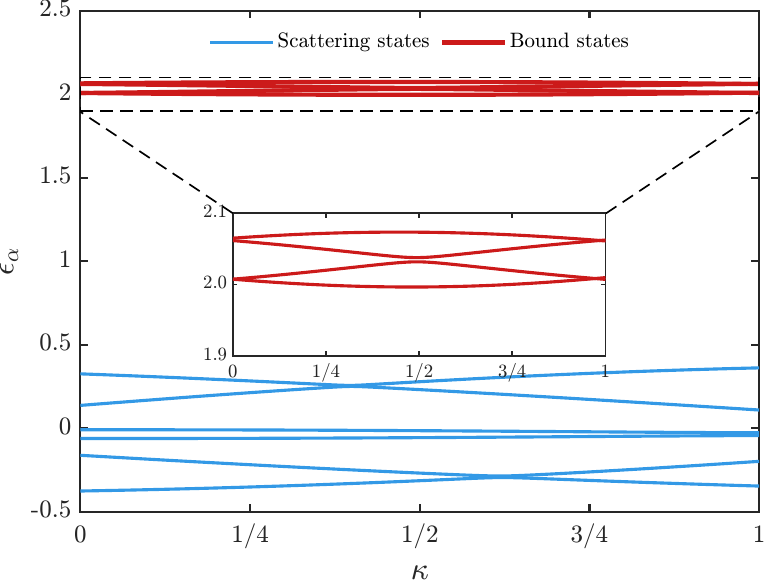} 
		\caption{Floquet quasi-energy spectrum of the two-anyon system governed by $\hat{H}_{\text{AHM}}^B(t)$ [Eqs.~\eqref{eq:full_hamiltonian} and \eqref{eq:scalar_potentials}] as a function of the statistical parameter $\kappa$. System and driving parameters are fixed at $U/J = 2$, $\omega/J = 40$, $Aa/\omega = 3.5$, and $\theta = \pi/8$. The high-frequency drive isolates the upper doublon (paired states) band (red) from the scattering (unpaired states) continuum (blue) even at a weak interaction strength of $U/J=2$. The inset resolves the fine structure of the doublon band, revealing exact twofold degeneracies at $\kappa = 0$ and $\kappa = 1$ corresponding to the dynamical caging conditions.}
		\label{fig:6}
	\end{figure}

	We turn to the quasi-energy spectrum to elucidate the physical origin of the Floquet-engineered caging effect. For a time-periodic system, the one-period dynamics are governed by the Floquet operator:
	\begin{equation} \label{eq:Floquet_operator}
		\hat{U}(T, 0) = \mathcal{T} \exp\left[-i \int_0^T \hat{H}_{\text{AHM}}^B(t) dt \right],
	\end{equation}
	where $\mathcal{T}$ denotes the time-ordering operator. According to Floquet's theorem, the eigenstates $|\phi_\alpha\rangle$ satisfy the eigenvalue equation $\hat{U}(T,0)|\phi_\alpha\rangle = e^{-i\varepsilon_\alpha T}|\phi_\alpha\rangle$. Here, $\alpha$ labels the distinct Floquet modes, and $\varepsilon_\alpha$ are the corresponding quasi-energies, which can be evaluated numerically by diagonalizing $\hat{U}(T,0)$. Figure~\ref{fig:6} presents the quasi-energy spectrum of the driven two-anyon system versus $\kappa$  at $\theta=\pi/8$  in the weakly interacting regime ($U/J = 2$). Thanks to the periodic driving, the spectrum is dynamically split into two widely separated bands: a broad lower scattering continuum (blue) and a narrow upper doublon (bound state) band (red) [Fig.~\ref{fig:6}]. Notably, as shown in the inset of Fig.~\ref{fig:6}, these doublon levels exhibit twofold degeneracies exactly at $\kappa = 0$ and $\kappa = 1$, directly coinciding with the onset of caging for this polarization angle. This implies that the Floquet-engineered caging effect is governed by the same physical mechanism as the statistically induced caging in the strongly interacting, undriven regime.
			
	To gain analytical insight into the Floquet-engineered caging effect under the resonance condition ($\Delta = m\omega$), we employ the high-frequency expansion method. We begin by introducing a time-dependent unitary transformation operator:
	\begin{equation} \label{eq:unitary_operator}
		\hat{\mathcal{R}}(t) = \exp\left[-i \sum_{j=1}^4 \chi_j(t) \hat{n}_j\right],
	\end{equation}
	where the accumulated phase is defined as $\chi_j(t) = \int_0^t V_j(t') dt'$. Transforming the system into the rotating frame yields the Hamiltonian $\hat{H}'(t)$:
	\begin{equation} \label{eq:H_transformed_merged}
		\begin{aligned}
			\hat{H}'(t) &= \hat{\mathcal{R}}^\dagger(t) \hat{H}_{\text{AHM}}^B(t) \hat{\mathcal{R}}(t) - i \hat{\mathcal{R}}^\dagger(t)\dot{\hat{\mathcal{R}}}(t) \\
			&= -J \sum_{j=1}^4 \left( \hat{b}_j^\dagger \hat{b}_{j+1} e^{-i\frac{\kappa\pi}{4}} e^{i\kappa\pi\hat{n}_j} e^{i[\chi_j(t) - \chi_{j+1}(t)]} + \text{H.c.} \right) \\
			&\quad + \frac{U}{2} \sum_{j=1}^4 \hat{n}_j(\hat{n}_j - 1).
		\end{aligned}
	\end{equation}
	
	In the high-frequency limit ($\omega \gg J, U$), the system's dynamics are accurately captured by a time-independent effective Floquet Hamiltonian, obtained by averaging $\hat{H}'(t)$ over one driving period $T$:
	\begin{equation} \label{eq:Heff_result}
		\begin{aligned}
			\hat{H}_{\text{eff}}^F &= \frac{1}{T} \int_0^T \hat{H}'(t) dt \\
			&= - \sum_{j=1}^4 \left( J_j^{\text{eff}} \hat{b}_j^\dagger \hat{b}_{j+1} e^{-i\frac{\kappa\pi}{4}} e^{i\kappa\pi\hat{n}_j} + \text{H.c.} \right) \\
			&\quad + \frac{U}{2} \sum_{j=1}^4 \hat{n}_j(\hat{n}_j - 1).
		\end{aligned}
	\end{equation}
	
	Integrating the oscillating phases in Eq.~\eqref{eq:H_transformed_merged} yields the renormalized single-particle hopping amplitudes: $J_1^{\text{eff}} = J_4^{\text{eff}} = J (-1)^m \mathcal{J}_m(Aa/\omega) e^{-im\theta}$ and $J_2^{\text{eff}} = J_3^{\text{eff}} = J \mathcal{J}_m(Aa/\omega) e^{-im\theta}$, where $\mathcal{J}_m$ is the $m$-th order Bessel function of the first kind. The high-frequency drive thereby renormalizes the hopping amplitudes to a uniform magnitude, $|J_j^{\text{eff}}| \equiv J_{\text{eff}} = J|\mathcal{J}_m(Aa/\omega)|$. Notably, at the roots of the Bessel function where $\mathcal{J}_m(Aa/\omega) = 0$, single-particle tunneling is completely suppressed, characterizing the coherent destruction of tunneling (CDT).

	In the regime $U \gg J_{\text{eff}}$—a condition dynamically achievable even for weak physical interactions $U$ by tuning the drive to suppress $J_{\text{eff}}$—second-order perturbation theory projects the effective Floquet Hamiltonian $\hat{H}_{\text{eff}}^F$ directly onto the doublon subspace $\{|2\rangle_j\}$. Following the exact derivation steps detailed in Sec.~\ref{sec:static}, we obtain the effective doublon Hamiltonian of the driven system:
	\begin{equation} \label{eq:driven_doublon_Heff}
		\begin{aligned}
			\hat{H}_{\text{eff}}^{F,D} &= \left( U + \frac{4J_{\text{eff}}^2}{U} \right) \sum_{j=1}^4 |2\rangle_j \langle 2|_j \\
			&\quad + \sum_{j=1}^4 \left( \frac{2J_{\text{eff}}^2}{U} e^{-i\phi^F} |2\rangle_{j+1} \langle 2|_j + \text{H.c.} \right),
		\end{aligned}
	\end{equation}
	where $\phi^F \equiv \kappa\pi/2 - 2m\theta$ defines the effective doublon phase incorporating the synthetic Flouqet flux. This effective Hamiltonian~\eqref{eq:driven_doublon_Heff} retains the identical algebraic structure of Eq.~\eqref{eq:H_eff_final}, simply replacing the static hopping amplitude $J$ with $J_{\text{eff}}$ and the statistical phase $\phi$ with $\phi^F$.

	\begin{figure}[t!]
	\centering
	\includegraphics[width=\linewidth]{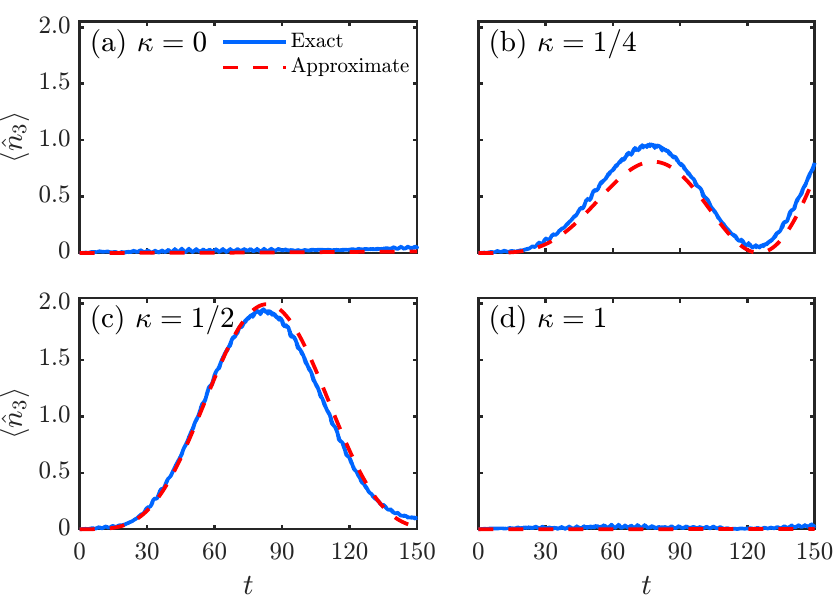} 
	\caption{Comparison of the target site occupation $\langle\hat{n}_3(t)\rangle$ governed by the full Hamiltonian $\hat{H}_{\text{AHM}}^B(t)$ [Eqs.~\eqref{eq:full_hamiltonian} and \eqref{eq:scalar_potentials}] (blue solid lines) and the effective  doublon  model $\hat{H}_{\text{eff}}^{F,D}$ [Eq.~\eqref{eq:driven_doublon_Heff}] (red dashed lines). The system and driving parameters are fixed at $U/J = 2$, $\omega/J = 40$, $Aa/\omega = 3.5$, and $\theta = \pi/8$. The dynamics are plotted for statistical parameters (a) $\kappa = 0$, (b) $\kappa = 1/4$, (c) $\kappa = 1/2$, and (d) $\kappa = 1$. The effective  doublon  model accurately captures the exact time evolution across all cases.}
	\label{fig:7}
	\end{figure}

	Next, building upon Eq.~\eqref{eq:driven_doublon_Heff}, we establish the caging condition for the driven system. Expanding the eigenstate as $|\psi\rangle = \sum_{j=1}^4 c_{j,j}|2\rangle_j$ and solving the eigenvalue equation $\hat{H}_{\text{eff}}^{F,D}|\psi\rangle = E|\psi\rangle$ yields a difference equation structurally identical to Eq.~\eqref{eq:difference_eq}. Eliminating the probability amplitudes of the even sites leads to a $2 \times 2$ secular equation for the odd sites:
	\begin{equation} \label{eq:driven_secular}
		\begin{cases}
			\left( \tilde{E}^2 - \frac{8J_{\text{eff}}^4}{U^2} \right) c_{1,1} - \frac{8J_{\text{eff}}^4}{U^2} \cos(2\phi^F) c_{3,3} = 0, \\
			- \frac{8J_{\text{eff}}^4}{U^2} \cos(2\phi^F) c_{1,1} + \left( \tilde{E}^2 - \frac{8J_{\text{eff}}^4}{U^2} \right) c_{3,3} = 0,
		\end{cases}
	\end{equation}
	where $\tilde{E} = E - (U + 4J_{\text{eff}}^2/U)$. Setting the off-diagonal coupling term to zero, $\cos(2\phi^F) = 0$, provides the exact condition for recovering the statistically induced caging:
	\begin{equation}
		8m\theta - 2\kappa\pi = (2n + 1)\pi, \quad (n \in \mathbb{Z}). \label{eq:driven_caging_condition}
	\end{equation}
	
	Under this condition, the synthetic flux penetrating the plaquette becomes $\Phi^F = 4\phi^F = -(2n + 1)\pi$. This demonstrates that the statistically induced caging is dynamically recovered when the drive-induced phase perfectly compensates for the anyonic statistical phase, thereby restoring the requisite $\pi$-flux condition for localization. Substituting the specific parameters $\theta = \pi/8$ and $m = 1$ into Eq.~\eqref{eq:driven_caging_condition} yields $2\kappa\pi = -2n\pi$. Evidently, the statistical parameters $\kappa = 0$ and $\kappa = 1$ simultaneously satisfy this caging condition. This analytical derivation conclusively confirms that the twofold band degeneracies at $\kappa = 0$ and $\kappa = 1$, as observed in Fig.~\ref{fig:6}, originate from exact destructive quantum interference, which is triggered when the Floquet driving phase and the anyonic statistical phase act synergistically to generate the requisite $\pi$-flux.

	\begin{figure}[t!]
		\centering
		\includegraphics[width=\linewidth]{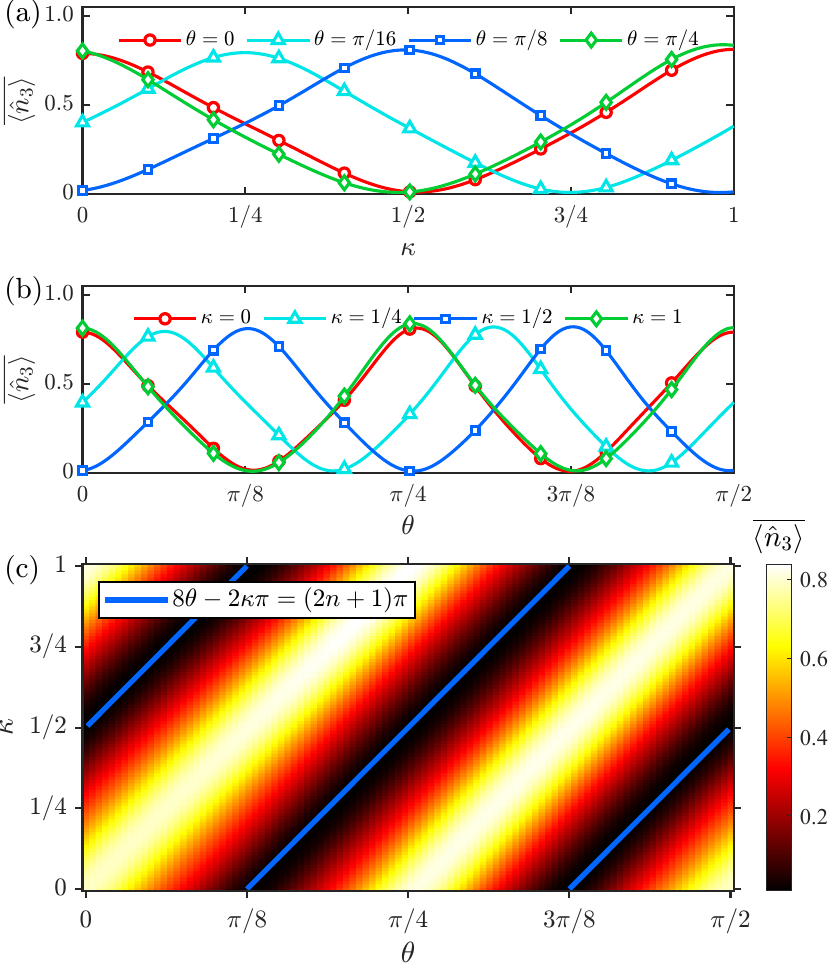} 
		\caption{Long-time-averaged target occupation $\overline{\langle \hat{n}_3 \rangle}$ governed by $\hat{H}_{\text{AHM}}^B(t)$ [Eqs.~\eqref{eq:full_hamiltonian} and \eqref{eq:scalar_potentials}] as a function of (a) the statistical parameter $\kappa$ for various polarization angles $\theta$, and (b) the polarization angle $\theta$ for various $\kappa$. (c) Two-dimensional color plot of $\overline{\langle \hat{n}_3 \rangle}$ in the $\theta$-$\kappa$ plane. Solid blue lines indicate the dynamical caging condition $8\theta - 2\kappa\pi = (2n+1)\pi$ derived from the effective model $\hat{H}_{\text{eff}}^{F,D}$ [Eq.~\eqref{eq:driven_doublon_Heff}]. Parameters are fixed at $U/J = 2$, $\omega/J = 40$, and $Aa/\omega = 3.5$.}
		\label{fig:8}
	\end{figure}

	To benchmark the validity of the effective Floquet doublon model, Fig.~\ref{fig:7} compares the time evolution of the target site occupation, $\langle\hat{n}_3(t)\rangle$, obtained via two independent approaches. We specifically examine the weakly interacting regime ($U/J = 2$) at a polarization angle of $\theta = \pi/8$ across four distinct statistical parameters ($\kappa = 0, 1/4, 1/2$, and $1$). The dynamics predicted by the effective doublon Hamiltonian [Eq.~\eqref{eq:driven_doublon_Heff}] are plotted against those obtained from numerical simulations of the full time-dependent Hamiltonian [Eq.~\eqref{eq:full_hamiltonian}, equivalent to Eq.~\eqref{eq:H_AHM}] with the potential in Eq.~\eqref{eq:scalar_potentials}. The results from both methods exhibit remarkable agreement. This not only confirms the accuracy of the high-frequency expansion but also unequivocally demonstrates that, for this specific polarization angle, $\theta = \pi/8$, the statistically induced caging effect emerges exclusively at $\kappa = 0$ and $\kappa = 1$.
	
	Leveraging this validated model, we can actively tune the caging of anyons based on the analytical condition in Eq.~\eqref{eq:driven_caging_condition}. As expected, this analytical prediction is fully corroborated by direct numerical simulations of the original time-periodic Hamiltonian $\hat{H}_{\text{AHM}}^B(t)$, as shown in Fig.~\ref{fig:8}. Figures~\ref{fig:8}(a) and \ref{fig:8}(b) display the dependence of the time-averaged target site occupation, $\overline{\langle\hat{n}_3\rangle}$, on the statistical parameter $\kappa$ and the polarization angle $\theta$, respectively, at $U/J=2$ and $Aa/\omega=3.5$. These numerical results clearly demonstrate that varying the polarization angle of the driving field can effectively induce caging (characterized by $\overline{\langle\hat{n}_3\rangle} \approx 0$) for anyons with distinct statistical parameters. In Fig.~\ref{fig:8}(c), we present the corresponding two-dimensional profile of $\overline{\langle\hat{n}_3\rangle}$ spanning the $\theta$-$\kappa$ plane. The numerically observed regions of zero occupation align perfectly with the analytical caging condition $8\theta - 2\kappa\pi = (2n + 1)\pi$, which is overlaid as solid blue lines for the single-photon resonance ($m = 1$). 
	
	These findings demonstrate that the synthetic Floquet flux and the anyonic statistical phase synergistically dictate the macroscopic quantum interference governing the caging effect. This fundamental interplay establishes a statistics-selective tuning mechanism: by merely adjusting the polarization angle $\theta$, one can selectively cage anyons characterized by specific statistical parameters, thereby providing a powerful dynamical pathway for probing and identifying anyonic statistics.

	\section{Conclusion}
	\label{sec:conclusion}
	
	In summary, we demonstrate statistical-factor-induced Aharonov–Bohm caging in the Anyon-Hubbard model on a four-site plaquette, a system exactly mappable to a Bose-Hubbard model. This caging, governed by synthetic Floquet flux and anyonic statistical phase, is highly tunable via Floquet engineering and accessible even in the weakly interacting regime---a regime inaccessible in the static case. By employing an effective Floquet Hamiltonian and second-order degenerate perturbation theory, we derive analytical caging conditions that show excellent agreement with direct numerical simulations of the original time-dependent system. Our results indicate that by tuning the polarization angle of the driving field, caging can be achieved for two interacting anyons with different statistical phases. This capability offers a versatile platform for quasiparticle manipulation, providing new insights for quantum simulation and the engineering of topological quantum matter in synthetic dimensions \cite{Goldman2016, Ozawa2019, Cooper2019}.
	
	\begin{acknowledgments}
		
	The work was supported by the National Natural Science Foundation of China (Grants No. 12375022, No. 11975110), and the Zhejiang Sci-Tech University Scientific Research Start-up Fund (Grant No. 20062318-Y). Z. Zhou and C. Guo contributed equally to this work.
	
	\end{acknowledgments}
	
	\bibliographystyle{apsrev4-2}
	\bibliography{refs}

@article{Leinaas1977,
	title   = {On the theory of identical particles},
	author  = {Leinaas, J. M. and Myrheim, J.},
	journal = {Nuovo Cimento B},
	volume  = {37},
	number  = {1},
	pages   = {1--23},
	year    = {1977},
	doi     = {10.1007/BF02727953}
}

@article{Wilczek1982,
	title   = {Quantum mechanics of fractional-spin particles},
	author  = {Wilczek, Frank},
	journal = {Phys. Rev. Lett.},
	volume  = {49},
	number  = {14},
	pages   = {957--959},
	year    = {1982},
	doi     = {10.1103/PhysRevLett.49.957}
}

@article{Halperin1984,
	title   = {Statistics of Quasiparticles and the Hierarchy of Fractional Quantized Hall States},
	author  = {Halperin, B. I.},
	journal = {Phys. Rev. Lett.},
	volume  = {52},
	number  = {18},
	pages   = {1583--1586},
	year    = {1984},
	doi     = {10.1103/PhysRevLett.52.1583}
}

@article{Arovas1984,
	title   = {Fractional Statistics and the Quantum Hall Effect},
	author  = {Arovas, Daniel and Schrieffer, J. R. and Wilczek, Frank},
	journal = {Phys. Rev. Lett.},
	volume  = {53},
	number  = {7},
	pages   = {722--723},
	year    = {1984},
	doi     = {10.1103/PhysRevLett.53.722}
}

@article{Stern2008,
	title   = {Anyons and the quantum Hall effect—A pedagogical review},
	author  = {Stern, Ady},
	journal = {Ann. Phys.},
	volume  = {323},
	number  = {1},
	pages   = {204--249},
	year    = {2008},
	doi     = {10.1016/j.aop.2007.10.008}
}

@article{Nayak2008,
	title   = {Non-Abelian anyons and topological quantum computation},
	author  = {Nayak, Chetan and Simon, Steven H. and Stern, Ady and Freedman, Michael and Das Sarma, Sankar},
	journal = {Rev. Mod. Phys.},
	volume  = {80},
	number  = {3},
	pages   = {1083--1159},
	year    = {2008},
	doi     = {10.1103/RevModPhys.80.1083}
}

@article{Feldman2021,
	title   = {Fractional charge and fractional statistics in the quantum Hall effects},
	author  = {Feldman, D. E. and Halperin, Bertrand I.},
	journal = {Rep. Prog. Phys.},
	volume  = {84},
	number  = {7},
	pages   = {076501},
	year    = {2021},
	doi     = {10.1088/1361-6633/ac03aa}
}

@article{Trung2021,
	title   = {Fractionalization and Dynamics of Anyons and Their Experimental Signatures in the $\nu=n+1/3$ Fractional Quantum Hall State},
	author  = {Trung, Ha Quang and Yang, Bo},
	journal = {Phys. Rev. Lett.},
	volume  = {127},
	number  = {4},
	pages   = {046402},
	year    = {2021},
	doi     = {10.1103/PhysRevLett.127.046402}
}

@article{Bartolomei2020,
	title   = {Fractional statistics in anyon collisions},
	author = {H. Bartolomei  and M. Kumar  and R. Bisognin  and A. Marguerite  and J.-M. Berroir  and E. Bocquillon  and B. Plaçais  and A. Cavanna  and Q. Dong  and U. Gennser  and Y. Jin  and G. Fève },
	journal = {Science},
	volume  = {368},
	number  = {6487},
	pages   = {173--177},
	year    = {2020},
	doi     = {10.1126/science.aaz5601}
}

@article{Nakamura2020,
	title   = {Direct observation of anyonic braiding statistics},
	author  = {Nakamura, J. and Liang, S. and Gardner, G. C. and Manfra, M. J.},
	journal = {Nat. Phys.},
	volume  = {16},
	number  = {9},
	pages   = {931--936},
	year    = {2020},
	doi     = {10.1038/s41567-020-1019-1}
}

@article{Google2023,
	title   = {Non-Abelian braiding of graph vertices in a superconducting processor},
	author  = {{Google Quantum AI and Collaborators}},
	journal = {Nature},
	volume  = {618},
	number  = {7964},
	pages   = {264--269},
	year    = {2023},
	doi     = {10.1038/s41586-023-05954-4}
}

@article{Strater2016,
	title   = {Floquet Realization and Signatures of One-Dimensional Anyons in an Optical Lattice},
	author  = {Str{\"a}ter, Christoph and Srivastava, Shashi C. L. and Eckardt, Andr{\'e}},
	journal = {Phys. Rev. Lett.},
	volume  = {117},
	number  = {20},
	pages   = {205303},
	year    = {2016},
	doi     = {10.1103/PhysRevLett.117.205303}
}

@article{Schweizer2019,
	title   = {Floquet approach to $\mathbb{Z}_2$ lattice gauge theories with ultracold atoms in optical lattices},
	author  = {Schweizer, Christian and Grusdt, Fabian and Berngruber, Moritz and Barbiero, Luca and Demler, Eugene and Goldman, Nathan and Bloch, Immanuel and Aidelsburger, Monika},
	journal = {Nat. Phys.},
	volume  = {15},
	number  = {11},
	pages   = {1168--1173},
	year    = {2019},
	doi     = {10.1038/s41567-019-0649-7}
}

@article{Kwan2024,
	title   = {Realization of one-dimensional anyons with arbitrary statistical phase},
	author  = {Kwan, Joyce and Segura, Perrin and Li, Yanfei and Kim, Sooshin and Gorshkov, Alexey V. and Eckardt, Andr{\'e} and Bakkali-Hassani, Brice and Greiner, Markus},
	journal = {Science},
	volume  = {386},
	number  = {6725},
	pages   = {1055--1060},
	year    = {2024},
	doi     = {10.1126/science.adi3252}
}

@article{Keilmann2011,
	title   = {Statistically induced phase transitions and anyons in 1D optical lattices},
	author  = {Keilmann, Tassilo and Lanzmich, Simon and McCulloch, Ian and Roncaglia, Marco},
	journal = {Nat. Commun.},
	volume  = {2},
	pages   = {361},
	year    = {2011},
	doi     = {10.1038/ncomms1353}
}

@article{Zhang2022,
	author = {Zhang, Weixuan and Yuan, Hao and Wang, Haiteng and Di, Fengxiao and Sun, Na and Zheng, Xingen and Sun, Houjun and Zhang, Xiangdong},
	title = {Observation of Bloch oscillations dominated by effective anyonic particle statistics},
	journal = {Nat. Commun.},
	volume = {13},
	pages = {2392},
	year = {2022},
	doi = {10.1038/s41467-022-29895-0}
}

@article{Longhi2012,
	title   = {Anyonic Bloch oscillations},
	author  = {Longhi, Stefano and Della Valle, Giuseppe},
	journal = {Phys. Rev. B},
	volume  = {85},
	number  = {16},
	pages   = {165144},
	year    = {2012},
	doi     = {10.1103/PhysRevB.85.165144}
}

@article{Greschner2015,
	title   = {Anyon Hubbard Model in One-Dimensional Optical Lattices},
	author  = {Greschner, S. and Santos, L.},
	journal = {Phys. Rev. Lett.},
	volume  = {115},
	number  = {5},
	pages   = {053002},
	year    = {2015},
	doi     = {10.1103/PhysRevLett.115.053002}
}

@article{Cardarelli2016,
	title   = {Engineering interactions and anyon statistics by multicolor lattice-depth modulations},
	author  = {Cardarelli, Lorenzo and Greschner, Sebastian and Santos, Luis},
	journal = {Phys. Rev. A},
	volume  = {94},
	number  = {2},
	pages   = {023615},
	year    = {2016},
	doi     = {10.1103/PhysRevA.94.023615}
}

@article{Zhang2023BIC,
	title   = {Anyonic bound states in the continuum},
	author  = {Zhang, Weixuan and Qian, Long and Sun, Houjun and Zhang, Xiangdong},
	journal = {Commun. Phys.},
	volume  = {6},
	number  = {1},
	pages   = {139},
	year    = {2023},
	doi     = {10.1038/s42005-023-01245-6}
}

@article{Qin2025,
	title   = {Dynamical suppression of many-body non-Hermitian skin effect in anyonic systems},
	author  = {Qin, Yi and Lee, Ching Hua and Li, Linhu},
	journal = {Commun. Phys.},
	volume  = {8},
	pages   = {18},
	year    = {2025},
	doi     = {10.1038/s42005-025-01935-3}
}

@article{Chamon1997,
	title   = {Two point-contact interferometer for quantum Hall systems},
	author  = {Chamon, C. de C. and Freed, D. E. and Kivelson, S. A. and Wen, X. G. and Vishwanath, A.},
	journal = {Phys. Rev. B},
	volume  = {55},
	number  = {4},
	pages   = {2331--2343},
	year    = {1997},
	doi     = {10.1103/PhysRevB.55.2331}
}

@article{Campagnano2012,
	title   = {Hanbury Brown-Twiss Interference of Anyons},
	author  = {Campagnano, G. and Zilberberg, O. and Gornyi, I. V. and Feldman, D. E. and Mirlin, A. D. and Gefen, Y.},
	journal = {Phys. Rev. Lett.},
	volume  = {109},
	number  = {10},
	pages   = {106802},
	year    = {2012},
	doi     = {10.1103/PhysRevLett.109.106802}
}

@article{Leykam2018,
	title   = {Artificial flat band systems: from lattice models to experiments},
	author  = {Leykam, Daniel and Andreanov, Alexei and Flach, Sergej},
	journal = {Adv. Phys. X},
	volume  = {3},
	number  = {1},
	pages   = {1473052},
	year    = {2018},
	doi     = {10.1080/23746149.2018.1473052}
}

@article{Vidal1998,
	title   = {Aharonov-Bohm caging in two-dimensional lattices},
	author  = {Vidal, Julien and Mosseri, R{\'e}my and Dou{\c{c}}ot, Beno{\^\i}t},
	journal = {Phys. Rev. Lett.},
	volume  = {81},
	number  = {26},
	pages   = {5888--5891},
	year    = {1998},
	doi     = {10.1103/PhysRevLett.81.5888}
}

@article{Mukherjee2018,
	title   = {Experimental observation of Aharonov-Bohm cages in photonic lattices},
	author  = {Mukherjee, S. and Di Liberto, M. and {\"O}hberg, P. and Thomson, R. R. and Goldman, N.},
	journal = {Phys. Rev. Lett.},
	volume  = {121},
	number  = {7},
	pages   = {075502},
	year    = {2018},
	doi     = {10.1103/PhysRevLett.121.075502}
}

@article{Li2022,
	title   = {Aharonov-Bohm Caging and Inverse Anderson Transition in Ultracold Atoms},
	author  = {Li, Hang and Dong, Zhaoli and Longhi, Stefano and Liang, Qian and Xie, Dizhou and Yan, Bo},
	journal = {Phys. Rev. Lett.},
	volume  = {129},
	number  = {22},
	pages   = {220403},
	year    = {2022},
	doi     = {10.1103/PhysRevLett.129.220403}
}

@article{Longhi2021,
	title   = {Inverse Anderson transition in photonic cages},
	author  = {Longhi, Stefano},
	journal = {Opt. Lett.},
	volume  = {46},
	number  = {12},
	pages   = {2872--2875},
	year    = {2021},
	doi     = {10.1364/OL.430196}
}

@article{Doucot2002,
	title   = {Pairing of Cooper Pairs in a Fully Frustrated Josephson-Junction Chain},
	author  = {Dou{\c{c}}ot, Beno{\^\i}t and Vidal, Julien},
	journal = {Phys. Rev. Lett.},
	volume  = {88},
	number  = {22},
	pages   = {227005},
	year    = {2002},
	doi     = {10.1103/PhysRevLett.88.227005}
}

@article{Creffield2010,
	title   = {Coherent Control of Interacting Particles Using Dynamical and Aharonov-Bohm Phases},
	author  = {Creffield, C. E. and Platero, G.},
	journal = {Phys. Rev. Lett.},
	volume  = {105},
	number  = {8},
	pages   = {086804},
	year    = {2010},
	doi     = {10.1103/PhysRevLett.105.086804}
}

@article{Zurita2020,
	title   = {Topology and interactions in the photonic Creutz and Creutz-Hubbard ladders},
	author  = {Zurita, Juan and Creffield, Charles E. and Platero, Gloria},
	journal = {Adv. Quantum Technol.},
	volume  = {3},
	number  = {2},
	pages   = {1900105},
	year    = {2020},
	doi     = {10.1002/qute.201900105}
}

@article{Vidal2000,
	title   = {Interaction induced delocalization for two particles in a periodic potential},
	author  = {Vidal, Julien and Dou{\c{c}}ot, Beno{\^\i}t and Mosseri, R{\'e}my and Pichard, Christophe},
	journal = {Phys. Rev. Lett.},
	volume  = {85},
	number  = {18},
	pages   = {3906--3909},
	year    = {2000},
	doi     = {10.1103/PhysRevLett.85.3906}
}

@article{Bukov2015,
	title   = {Universal high-frequency behavior of periodically driven systems: from dynamical stabilization to Floquet engineering},
	author  = {Bukov, Marin and D'Alessio, Luca and Polkovnikov, Anatoli},
	journal = {Adv. Phys.},
	volume  = {64},
	number  = {2},
	pages   = {139--226},
	year    = {2015},
	doi     = {10.1080/00018732.2015.1055918}
}

@article{Eckardt2017,
	title   = {Colloquium: Atomic quantum gases in periodically driven optical lattices},
	author  = {Eckardt, Andr{\'e}},
	journal = {Rev. Mod. Phys.},
	volume  = {89},
	number  = {1},
	pages   = {011004},
	year    = {2017},
	doi     = {10.1103/RevModPhys.89.011004}
}

@article{Goldman2014,
	title   = {Periodically driven quantum systems: Effective Hamiltonians and engineered gauge fields},
	author  = {Goldman, N. and Dalibard, J.},
	journal = {Phys. Rev. X},
	volume  = {4},
	number  = {3},
	pages   = {031027},
	year    = {2014},
	doi     = {10.1103/PhysRevX.4.031027}
}

@article{Eckardt2015,
	title   = {High-frequency approximation for periodically driven quantum systems from a Floquet-space perspective},
	author  = {Eckardt, Andr{\'e} and Anisimovas, Egidijus},
	journal = {New J. Phys.},
	volume  = {17},
	number  = {9},
	pages   = {093039},
	year    = {2015},
	doi     = {10.1088/1367-2630/17/9/093039}
}

@article{Luo2021,
	title   = {Controlling directed atomic motion and second-order tunneling of a spin-orbit-coupled atom in optical lattices},
	author = {Luo, Xiaobing and Zeng, Zhao-Yun and Guo, Yu and Yang, Baiyuan and Xiao, Jinpeng and Li, Lei and Kong, Chao and Chen, Ai-Xi},
	journal = {Phys. Rev. A},
	volume  = {103},
	number  = {4},
	pages   = {043315},
	year    = {2021},
	doi     = {10.1103/PhysRevA.103.043315}
}

@article{Wu2024,
	title   = {Spin-orbit coupling effects on localization and correlated tunneling for two interacting bosons in a double-well potential},
	author = {Wu, Hongzheng and Yan, Xin and Fan, Changwei and Yang, Baiyuan and Xiao, Jinpeng and Zeng, Zhao-Yun and Chen, Yajiang and Luo, Xiaobing},
	journal = {New J. Phys.},
	volume  = {26},
	number  = {4},
	pages   = {043020},
	year    = {2024},
	doi     = {10.1088/1367-2630/ad3be3}
}

@article{Struck2012,
	title   = {Tunable gauge potential for neutral and spinless particles in driven optical lattices},
	author  = {Struck, J. and {\"O}lschl{\"a}ger, C. and Weinberg, M. and Hauke, P. and Simonet, J. and Eckardt, A. and Lewenstein, M. and Sengstock, K. and Windpassinger, P.},
	journal = {Phys. Rev. Lett.},
	volume  = {108},
	number  = {22},
	pages   = {225304},
	year    = {2012},
	doi     = {10.1103/PhysRevLett.108.225304}
}

@article{Jotzu2014,
	title   = {Experimental realization of the topological Haldane model with ultracold fermions},
	author  = {Jotzu, Gregor and Messer, Michael and Desbuquois, R{\'e}mi and Lebrat, Martin and Uehlinger, Thomas and Greiner, Daniel and Esslinger, Tilman},
	journal = {Nature},
	volume  = {515},
	number  = {7526},
	pages   = {237--240},
	year    = {2014},
	doi     = {10.1038/nature13915}
}

@article{Grossmann1991,
	title   = {Coherent destruction of tunneling},
	author  = {Grossmann, F. and Dittrich, T. and Jung, P. and H{\"a}nggi, P.},
	journal = {Phys. Rev. Lett.},
	volume  = {67},
	number  = {4},
	pages   = {516--519},
	year    = {1991},
	doi     = {10.1103/PhysRevLett.67.516}
}

@article{Lignier2007,
	title   = {Dynamical Control of Matter-Wave Tunneling in Periodic Potentials},
	author  = {Lignier, H. and Sias, C. and Ciampini, D. and Singh, Y. and Zenesini, A. and Morsch, O. and Arimondo, E.},
	journal = {Phys. Rev. Lett.},
	volume  = {99},
	number  = {22},
	pages   = {220403},
	year    = {2007},
	doi     = {10.1103/PhysRevLett.99.220403}
}

@article{Eckardt2009,
	title   = {Exploring dynamic localization with a Bose-Einstein condensate},
	author  = {Eckardt, Andr{\'e} and Holthaus, Martin and Lignier, Hans and Zenesini, Alessandro and Ciampini, Donatella and Morsch, Oliver and Arimondo, Ennio},
	journal = {Phys. Rev. A},
	volume  = {79},
	number  = {1},
	pages   = {013611},
	year    = {2009},
	doi     = {10.1103/PhysRevA.79.013611}
}

@article{Zenesini2009,
	title   = {Coherent Control of Dressed Matter Waves},
	author  = {Zenesini, Alessandro and Lignier, Hans and Ciampini, Donatella and Morsch, Oliver and Arimondo, Ennio},
	journal = {Phys. Rev. Lett.},
	volume  = {102},
	number  = {10},
	pages   = {100403},
	year    = {2009},
	doi     = {10.1103/PhysRevLett.102.100403}
}

@article{Greschner2014,
	title   = {Density-Dependent Synthetic Gauge Fields Using Periodically Modulated Interactions},
	author  = {Greschner, S. and Sun, G. and Poletti, D. and Santos, L.},
	journal = {Phys. Rev. Lett.},
	volume  = {113},
	number  = {21},
	pages   = {215303},
	year    = {2014},
	doi     = {10.1103/PhysRevLett.113.215303}
}

@article{Goldman2016,
	author  = {Goldman, N. and Budich, J. and Zoller, P.},
	title   = {Topological quantum matter with ultracold gases in optical lattices},
	journal = {Nat. Phys.},
	volume  = {12},
	pages   = {639--645},
	year    = {2016},
	doi     = {10.1038/nphys3803}
}

@article{Ozawa2019,
	title     = {Topological photonics},
	author    = {Ozawa, Tomoki and Price, Hannah M. and Amo, Alberto and Goldman, Nathan and Hafezi, Mohammad and Lu, Ling and Rechtsman, Mikael C. and Schuster, David and Simon, Jonathan and Zilberberg, Oded and Carusotto, Iacopo},
	journal   = {Rev. Mod. Phys.},
	volume    = {91},
	issue     = {1},
	pages     = {015006},
	numpages  = {76},
	year      = {2019},
	month     = {Mar},
	publisher = {American Physical Society},
	doi       = {10.1103/RevModPhys.91.015006}
}

@article{Cooper2019,
	title     = {Topological bands for ultracold atoms},
	author    = {Cooper, N. R. and Dalibard, J. and Spielman, I. B.},
	journal   = {Rev. Mod. Phys.},
	volume    = {91},
	issue     = {1},
	pages     = {015005},
	numpages  = {55},
	year      = {2019},
	month     = {Mar},
	publisher = {American Physical Society},
	doi       = {10.1103/RevModPhys.91.015005}
}
\end{document}